# Review on zero-resistance phenomena in mixed phases

Ji-Hai Yuan[1,2], Ya-Dong Gu[1,2], Qing-Song Liu[1,2], Jun-Kun Yi[1,2], Hao-Yu He[1,2], Meng Wang[1,2], Jia-Sheng Liu[1,2], Zheng-Xin Lin[1,2], Zhi-An Ren[1,2*]

[1] *Institute of Physics and Beijing National Laboratory for Condensed Matter Physics, Chinese Academy of Sciences, Beijing 100190, China*

[2] *School of Physical Sciences, University of Chinese Academy of Sciences, Beijing 100049, China*

[*] Corresponding author. E-mail: renzhian@iphy.ac.cn

Zero-resistance phenomena are a characteristic feature of superconductors. The four-probe (or van der Pauw) method is the most widely used technique for measuring small resistance in laboratory, by separating current injection from voltage detection and eliminating lead and contact resistance errors. In 2023, Hirsch showed through simulations that decoupled voltage probes in four-probe measurements on inhomogeneous samples can yield spurious resistance drops [1]. Later we observed this type of false zero-resistance transition in mixed phase samples experimentally [2]. In history, a number of unsolved zero-resistance phenomena have been reported. Here we briefly review the zero-resistance transitions in $GaNb_4Se_8/NbSe_2$ composites, some high-$T_c$ materials of unclear origin, and superhydride superconductors, and discuss their similarities.

## 1. The false zero-resistance transitions in $GaNb_4Se_8/NbSe_2$ mixed-phase samples

The Ge-doped $Ga_{1-x}Ge_xNb_4Se_8$ ($0 < x < 0.2$) polycrystalline samples were synthesized by solid-state reaction method with high-purity (99.99%) Ga, Nb, Se, Ge ingredients. The final products are black in color and stable in air. Powder XRD analysis indicates that the samples are mainly composed of two distinguishable chemical phases, the Mott insulating $GaNb_4Se_8$ and the metallic $NbSe_2$.

The temperature dependence of electrical resistance for three representative samples with resistance transitions, measured using the four-probe method, is shown in Fig. 1a-c. A sharp resistive transition at 72 K is observed for the sample displayed in Fig. 1a. Below 20 K, the measured resistance is on the order of $10^{-5}$ ohm, corresponding to a voltage of approximately $10^{-8}$ V under an applied current of 1 mA, approaching the lowest detection limit of the PPMS measurement system. We note that the observed resistive transition and the emergence of zero resistance at low temperature are almost indistinguishable from typical superconducting transitions. As additional samples were measured, we observed that some samples do not reach zero resistance at low temperatures but instead exhibit a residual resistance below the transition. The residual resistances at low temperatures vary randomly from sample to sample, taking on positive or even negative values. For instance, Fig. 1b and Fig. 1c present two samples displaying positive and negative residual resistance, respectively. The onset transition temperatures range between 50 K and 80 K. Magnetic susceptibility measurements on all these samples revealed no diamagnetic signal indicative of superconductivity.

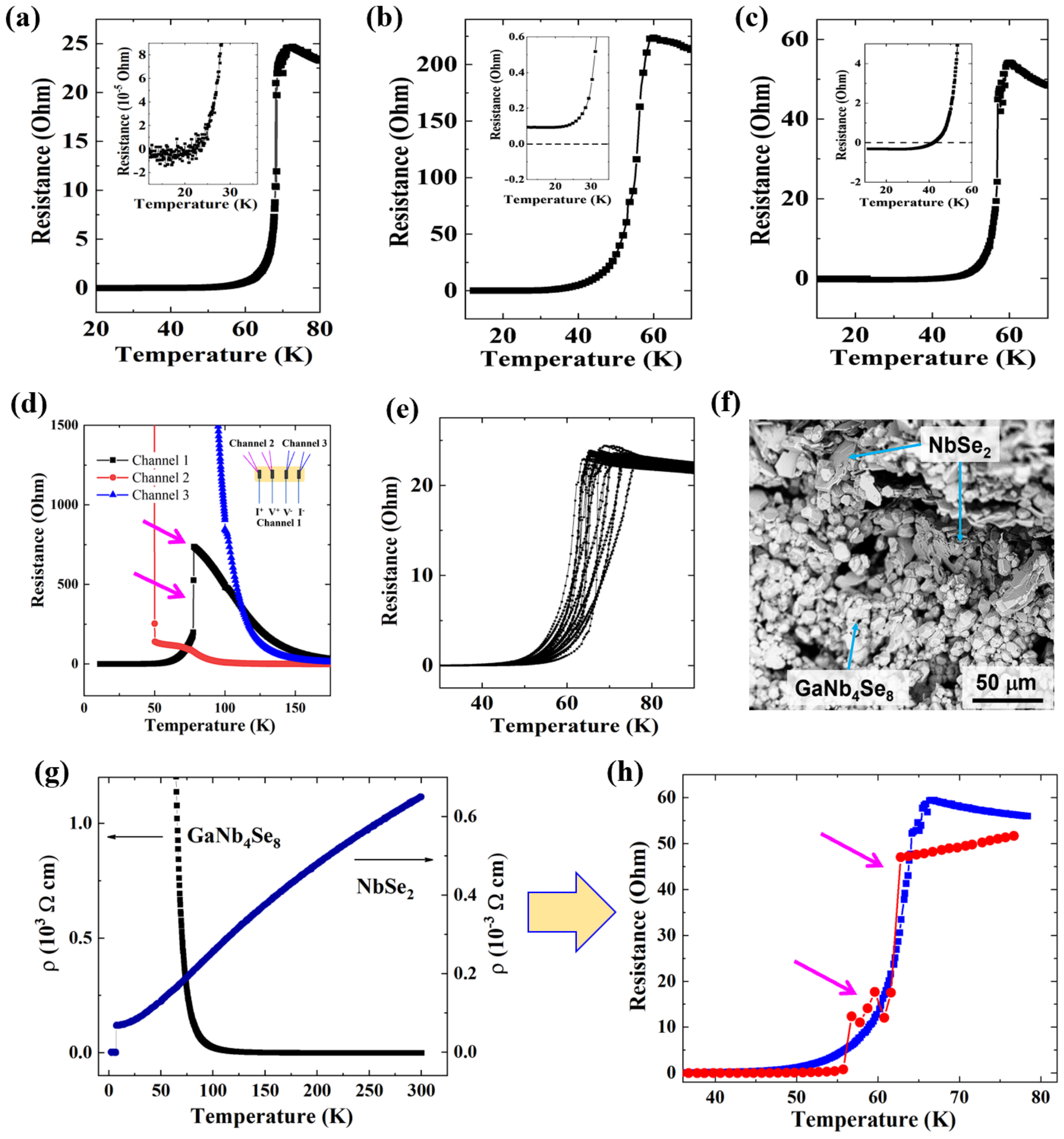

Figure 1. (a, b, c) Temperature dependence of the electrical resistances for three nominal $Ga_{1-x}Ge_xNb_4Se_8$ samples showing the false zero-resistance transitions. The inset shows the residual resistances below the transition temperatures. (d) Measurement of the temperature dependence of the electrical resistance on a sample with three channels. (e) A sample's resistive transition shifting under magnetic fields irregularly. (f) A SEM image showing the $GaNb_4Se_8$ and $NbSe_2$ grains in the sample. (g, h) Resistivity measurements for phase-pure $GaNb_4Se_8$ and $NbSe_2$, and false zero-resistance transitions in their mixed phases.

To understand the origin of these zero-resistance transitions, we devised an alternative measurement scheme. As illustrated in the inset of Fig. 1d, the three measurement channels of the PPMS were employed to probe the resistances of the three segments of the sample between the four contact leads. The PPMS performed sequential continuous measurements of resistance across all channels without mutual interference. As shown in Fig. 1d, when a sharp resistive transition occurs, the resistance of channel 3 becomes extremely large due to the insulating state. The $V^-$ lead is then isolated from the current and no current flows in the sample between the two voltage leads. At lower temperatures, the current flows only through the current leads between the $I^+$ and $I^-$ contacts, whereas the voltage leads ($V^+$ and $V^-$) both became electrically isolated. Therefore, the observed zero-resistance is a spurious phenomenon and does not reflect the intrinsic resistive characteristic of the sample.

We note that this false zero-resistance transition shifts with applied magnetic field clearly as shown in the inset of Fig. 1e for fields up to16 T. It is also sensitive to the measurement current. However, the shifting direction varies irregularly sample by sample, and for the same sample the behavior can also change over time. Moreover, these transition curves can cross each other under different fields. In some samples, the transition exhibits thermal hysteresis between heating and cooling, and the transition temperature also drifts over time. In contrast, other samples exhibit relatively stable transition temperatures for months. Taken together, the observations reveal no systematic pattern for these samples.

A Scanning electron microscopy image in Fig. 1f displays the different crystal grains of octahedral-like $GaNb_4Se_8$ and plate-like $NbSe_2$ in one sample. For reasons that remain unclear, in some samples, $NbSe_2$ grains formed conducting channels spontaneously rather than being randomly dispersed when sample was synthesized. In Fig. 1g, the resistivity measurements for pure $GaNb_4Se_8$ and $NbSe_2$ are displayed, which exhibit insulating and metallic behaviors as a function of temperature. In their mixed phase samples, typical false zero-resistance transitions can occur as in Fig. 1h.

For some samples, unlike the common resistive transition of multiphase polycrystalline superconductors which have broad resistive transitions originating from the spread of critical temperatures among individual grains and weak-link behavior at grain boundaries, some distinctive features are observed, such as *abrupt onset, sharp resistive drops, oscillations in the transition*, as indicated by the arrows in Fig. 1d, 1h. Clear magnetoresistance above the transition also exhibits in some samples. The oscillations in resistive transition can be understood as follows. When one of the voltage leads becomes isolating, the nearby current approaches zero. For inhomogeneous samples carrying very low current, switching between distinct current pathways may occur due to unstable internal circumstances, giving rise to the occasional oscillations in resistivity.

In 2023, using two-dimensional resistor-network simulations for the van der Pauw four-probe measurement, Hirsch presented that when portions of an inhomogeneous sample turn semiconducting or insulating with cooling, the voltage-probe-covered region may become effectively decoupled from current-carrying paths, yielding artificially tiny measured voltages and apparent resistance drops. Step-like features and oscillations in the resistive transition commonly observed in hydride transport measurements, can also arise from random connection-disconnection events among conducting pathways for different semiconducting resistors in the simulations [1].

Across all the $Ga_{1-x}Ge_xNb_4Se_8$ samples we synthesized, this false zero-resistance phenomenon was observed with a probability of 3%. This phenomenon is reproducible and is not restricted to a single sample. Because this phenomenon requires only the coexistence of metallic and insulating phases and it is detected with the widely used four-probe method in resistive measurements, it may also have been observed in previous experiments. We note that different sample and probe geometries may lead to different success rates, and high rate may occur in two-dimensional (2D) cases. In the following, we review several materials in which high-$T_c$ zero-resistance phenomena have been reported previously.

## 2. Six classes of high-$T_c$ zero-resistance materials

### 1) The doped lead apatite $Pb_{10-x}Cu_x(PO_4)_6O$

In July 2023, Lee *et al*. claimed room-temperature superconductivity below 127 $^{o}$C in the Cu-doped lead apatite $Pb_{10-x}Cu_x(PO_4)_6O$ (LK-99, Fig. 2a) [3]. Numerous subsequent replication attempts refuted this extraordinary claim and finally attributed the resistive signatures to $Cu_2S$ impurities, whose first-order structural transition near 400 K leads to the sudden drop in resistivity and mimics a superconducting transition without reaching zero resistance [4].

In the widespread effort to replicate room-temperature superconductivity, Hou *et al*. reported zero-resistance in $Pb_{10-x}Cu_x(PO_4)_6O$ samples by four-probe method, with the resistive transition shifting to lower temperatures under applied magnetic fields (Fig. 2b) [5]. Clear magnetoresistance above the transition exhibits in some samples. No Meissner effect was observed. The authors ruled out superconductivity and proposed a current-percolation model based on the formation of a Cu/Pb current channel to account for the observed zero-resistance behavior. Notably, although the $Pb_{10-x}Cu_x(PO_4)_6O$ samples are semiconducting, they often exhibit metallic behavior arising from impurities.

The LK-99 episode further highlights the limited reliability of magnetic susceptibility measurements in identifying superconductivity: different research groups reported paramagnetism, weak diamagnetism, ferromagnetism, spin-glass behavior, full diamagnetism, and even half-levitation for their replicated samples.

### 2) $Ca_2RuO_4$ nanofilms with $T_c$ ~ 96 K

In 2017, Nobukane *et al*. claimed superconductivity with an onset $T_c$ near 96 K in the Mott insulator $Ca_2RuO_4$ upon reducing the film thickness to a nanometer scale, and zero-resistance was observed in electrical transport measurements on these nanofilms, coexisting with ferromagnetism [6]. Furthermore, the authors reported that the temperature dependence of the resistivity was not suppressed under a magnetic field applied parallel to the c axis, and that the superconducting critical current exhibited a logarithmic temperature dependence and was enhanced by an external magnetic field, both of which are inconsistent with common superconductors. Moreover, the authors found that the induced bias current and the tuned film thickness may cause a superconductor-insulator transition. The resistivity values reported for nanofilms are also highly inconsistent, varying near three orders of magnitude at 300 K, and oscillations in the resistive transitions were also observed (Fig. 2c).

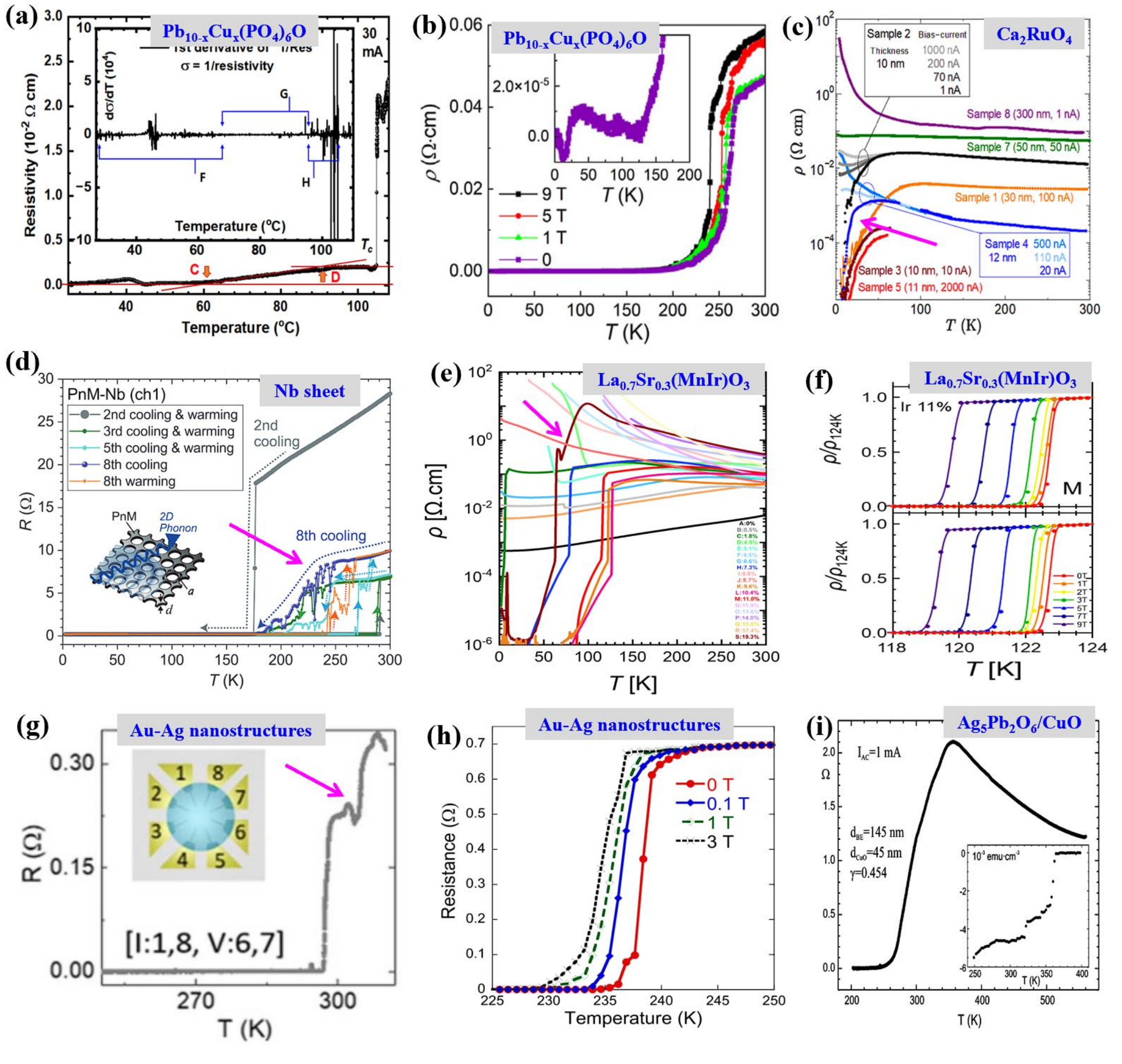

Figure 2. (a) Temperature dependence of resistivity for $Pb_{10-x}Cu_x(PO_4)_6O$ in which room temperature superconductivity was claimed [3]. (b) Resistive transitions for $Pb_{10-x}Cu_x(PO_4)_6O$ under magnetic field [5]. (c) Resistive transitions for $Ca_2RuO_4$ crystals [6]. (d) Resistive transitions of the Nb sheet sample [7]. (e, f) Resistive transitions of Ir-substituted $La_{0.7}Sr_{0.3}(MnIr)O_3$ films and magnetic-field response [8]. (g, h) Resistive transitions in Ag-Au nanostructures and field response [9]. (i) Resistive transition for $Ag_5Pb_2O_6$/CuO composite [10].

**3) Nb sheet with $T_c$ ~ 290 K**

Zen claimed that suspended Nb films with periodic through-holes enter a zero-resistance state superconductivity with an onset $T_c$ near 290 K Since 2020 [7]. The periodic holes were intended to fold the acoustic phonon spectrum via Bragg interference as a phonon-engineering approach. The Nb sheet is partially oxidized, forming a niobium-oxygen square lattice. Early reports noted that some samples exhibited Mott-insulating behavior. The high-$T_c$ resistive transitions are extremely sharp; large thermal hysteresis and oscillations in the resistive transition were also observed (Fig. 2d). The author also reported a weak magnetic drop and flux trapping at 300 K. To date, 7 of 18 rectangular Nb film samples examined have exhibited zero resistance transitions at high temperatures, corresponding to a success rate of 39%.

**4)** $La_{0.7}Sr_{0.3}(MnIr)O_3$ **with $T_c$ ~ 123 K**

In 2022, Matsumoto *et al.* reported superconductivity with an onset $T_c$ of 123 K and a zero-resistance state in Ir-substituted $La_{0.7}Sr_{0.3}(MnIr)O_3$ epitaxial films (Fig. 2e), motivated by the search for novel high-$T_c$ superconductors among perovskite oxides beyond cuprates. Weak ferromagnetic order coexists with superconductivity in these samples. The resistive transition temperature decreased steadily with increasing in-plane or out-of-plane magnetic field (Fig. 2f), as expected for a genuine superconductor. Although XPS scans claimed the absence of impurities in all film samples, the samples appear heterogeneous probably because of the difficulty of achieving high doping concentrations, and exhibiting metallic or semiconducting behaviors in resistive curves. The resistive transitions display sharp drops and occasional oscillations [8].

**5) Au-Ag nanostructures with $T_c$ ~ 298 K**

Since 2018, Thapa *et al.* reported ambient temperature superconductivity with a resistive transition $T_c$ ~ 298 K by four-probe measurement in films and pellets of a nanostructured material that is composed of silver particles embedded into a gold matrix (Fig. 2g), motivated by an intention to explore plasmon mediated electron-electron interaction effects. Later this zero-resistance transition was also reproduced by other research groups. Diamagnetic transition was reported in some samples. The $T_c$ was depended on the sample preparation and processing history. In cases, the state was seen to persist beyond measurement range (350 K - 400 K depending on the experiment). The nanostructures are sensitive to air and degrade over time. The resistive transition temperature was observed to shift to lower temperatures under applied magnetic fields (Fig. 2h). Very sharp transitions, large thermal hysteresis and oscillations in the resistive transitions were observed in some samples. Large spatial inhomogeneity has been reported in these thin films, with samples exhibiting either metallic or semiconducting behavior [9].

**6) $Ag_5Pb_2O_6$/CuO composite with $T_c$ ~ 356 K**

Djurek claimed that the $Ag_5Pb_2O_6$/CuO composite exhibits properties characteristic of room temperature superconductivity with a $T_c$ ranging from 178 K to 356 K, and a clear zero-resistance state with a possible one-dimensional conduction mechanism (Fig. 2i). This is a long research story spanning more than two decades by the same author Since 1990. The composite was reported to consist of metallic $Ag_5Pb_2O_6$ particles dispersed in a powdered insulating CuO matrix. Depending on the metallic-particle concentration, the mixture exhibits semiconducting or metallic behavior. This composite has been investigated extensively, and sharp or broad resistive transitions with zero-resistance have been observed in plenty of measurements. Weak diamagnetic transitions associated with the resistance transition were reported in several samples. However, no additional strong evidence supporting superconductivity has been reported [10].

In all these six classes of materials, zero-resistance state has been carefully confirmed by multiple measurements, with the measured resistance reaching the lower detection limit of the instruments. However, genuine superconductivity has not been validated in any of them. The question then arises: if these zero-resistance states are not superconducting, what is their origin?

In addition to occurring in thousands of known superconductors, a zero-resistance state has also been observed in quantum Hall materials and in electromagnetic-wave-excited GaAs/AlGaAs heterostructures under highly controlled experimental conditions [11-12]. Pure noble metals with resistivity below $10^{-9}$ ohm·m and millimeter sizes can also exhibit a resistance below the measurement resolution in laboratory. There are also measurement artifacts that can produce a zero-resistance reading, typically arising from wiring shorts, open circuits, high-impedance inputs, thermal EMFs, or sample breakage, *etc*. Such artifacts can be easily excluded by post-measurement inspections or repeated measurements. In samples containing precipitated metallic phases, like $Pb_{10-x}Cu_x(PO_4)_6O$ or $Ag_5Pb_2O_6$/CuO, current percolation or aggregation of metallic particles by thermal expansion and contraction has been proposed to explain the zero-resistance transitions. However, the migration and aggregation of metallic particles in a solid are unlikely to occur at low temperatures, and these scenarios would also yield a non-negligible residual resistance because of the poor connectivity between the metallic particles.

A plausible explanation is that the phenomena described above share the same false zero-resistance mechanism. All of them are mixed-phase systems containing both metallic and insulating phases, and the zero-resistance transitions were observed by four-probe method measurements. In such systems, the current path can be dominated by metallic phase while the voltage probes being decoupled from the current by insulating phases, thereby producing a spurious zero-resistance transition when temperature decreases. The presence of the insulating phase is essential, which leads to a zero-voltage reading in measurements by isolating the current from the voltage probes when its resistivity becomes high enough at certain low temperature that may be misidentified as the critical temperature $T_c$.

*Statistically, an event with non-zero probability will, given the necessary conditions, occur almost surely in repeated trials.*

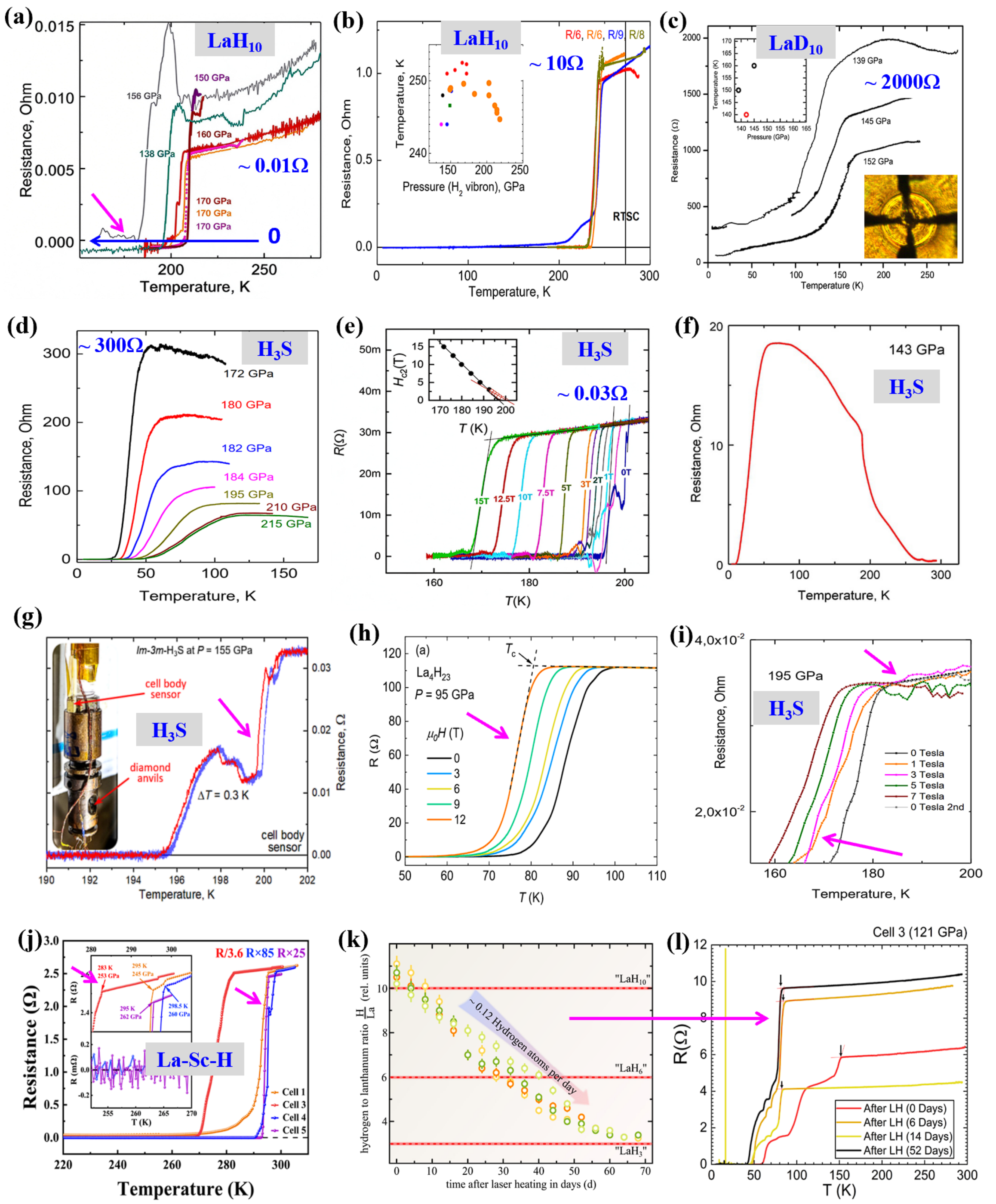

Figure 3. (a, b, c) Temperature dependence of resistance for $LaH_{10}$/$LaD_{10}$ superhydrides [17]. (d, e, f) Temperature dependence of resistance for $H_3S$ superhydrides [13]. (g) Oscillations in the resistive transition of $H_3S$ [13]. (h) Resistive transitions for $La_4H_{23}$ under magnetic field [19]. (i) Resistive transitions for $H_3S$ under magnetic field [13]. (j) Room-temperature superconductivity at 298 K in La-Sc-H system [14]. (k, l) Long-term evolution of hydrogen contents in $LaH_{10}$ and resistive transition measurements [21].

## 3. Anomalies in resistive transitions of superhydrides

Since the first experimental report in 2014 of superconductivity at 190 K in $H_3S$ by Drozdov *et al.*, several dozen hydride superconductors have been discovered to date [13]. In 2025, room-temperature superconductivity at 298 K was reported in the ternary La-Sc-H system [14]. Despite the evidence advanced in support of hydride superconductivity, substantial skepticism remains, and many anomalies persist in the reported data.

### 1) Sample synthesis of superhydrides

Superhydrides are generally synthesized at megabar pressures in diamond anvil cells by reacting pure elements (metals or sulfur, or their stable hydrides such as $LaH_3$) with hydrogen sources such as hydrogen gas, ammonia borane, or hydrocarbons under scanning pulsed-laser heating. Some authors noted that small regions of the sample are not fully heated to prevent laser damage to the pre-positioned electrical leads. From a materials synthesis perspective, heterogeneity is unavoidable in this synthesis process because of nonstoichiometry and nonuniform heating. Near the electrodes, this heterogeneity may be unintentionally enhanced.

Synchrotron X-ray diffraction measurements were used for phase identification and structural characterization. Spatially resolved XRD imaging was performed by raster scanning with a focused synchrotron beam (~1μm) to map the phase distribution. In almost all reported datasets, the samples are always inhomogeneous and exhibit unidentified diffraction peaks. The impurity phases have not been studied in detail. It is known that $LaH_3$ is a semiconducting phase, and no superconductivity has been reported for it. A study on experimental XRD data exhibits nanosized grains with average size of ~ 24 nm [15].

Balakirev *et al.* reported an overall success rate of about 25% for synthesizing the final high-temperature phases (approximately 300 trials in total), with all successfully synthesized samples exhibiting superconducting transitions. The unsuccessful attempts were attributed to the complexity and difficulty of the high-pressure experiments [16].

### 2) Large sample-to-sample variation in normal-state resistivity

Above the resistive transition, most superhydrides exhibit metallic behavior in the normal state, while some exhibits semiconducting behavior (Fig. 3d, 3f). In the normal state near 300 K, where the resistivity is dominated by electron-phonon scattering, many hydride samples exhibit large sample-to-sample variations. For $LaH_{10}$, the resistance reported by the same lab varies from 0.01 to 10 ohm and 2000 ohm (Fig. 3a-c) [17]. For $H_3S$, it varies from 0.03 to 300 ohm (Fig. 3d, 3e) [13]. For $LaSc_2H_{24}$, it varies from 0.05 to 200 ohm (Fig. 3j) [14]. This is far beyond the sample size changes for calculating resistivity. This large variation actually reflects the existing of insulating phases, and the changes of different chemical phases that dominate the electrical conduction in the mixture. However, although their resistances differ by a factor of several thousand, many samples exhibit similar sharp superconducting transitions with zero resistance at similar $T_c$ by experimental reports.

### 3) Unusual resistive transitions

Many of the superhydrides exhibit very narrow resistive transitions unlike typical inhomogeneous superconductors, which has been discussed by Hirsch previously [18]. There are more anomalies in the resistive transitions of hydrides, including the abrupt onset, sharp resistive drops, and oscillations in the transition, as indicated in Fig. 3 by arrows. Particularly, the specific oscillations in the transition, that appeared in many hydrides, are very similar with transitions in Fig. 2. In $La_4H_{23}$, an anomalous resistive transition is observed under an applied magnetic field, and the transition becomes narrower as the field increases [19]. Fig. 3i shows the irregular resistive transitions under an applied magnetic field for $H_3S$, including intersecting curves and obvious magnetoresistance above the resistive transition.

**4) Residual resistance below the transition**

Below the resistance transitions, positive residual resistances are often observed in many reports. This is explained by the presence of non-superconducting phases in the mixed-phase samples. There are several reports that the residual resistances exhibit negative values. Typically, as shown in Fig. 3a, 6 of the 7 measurements exhibit negative residual resistance, while one exhibits positive residual resistance [17]. This negative resistance is sufficiently large that one should be very cautious to explain it in terms of thermal EMFs, which are usually tiny small values that can often be neglected in common laboratory.

**5) Independent replication experiments**

Supporting evidence for superconductivity in superhydrides is that some of the zero-resistance transitions have been replicated by independent verifications. Although different $T_c$ values are reported by most groups for the same composition, these discrepancies are attributed to different superhydride phases or hydrogen content variation. Here we review two reports. Matsumoto *et al*. fabricated boron-doped diamond microelectrodes to improve electrical transport measurements on $H_2S$, which transforms into the superconductor $H_3S$ under high pressure. The measurements were conducted under high pressure up to 192 GPa but showed no evidence of high-$T_c$ transitions [20]. Zhou *et al*. performed combined $^{1}H$- and $^{139}La$-NMR measurements on lanthanum superhydrides, $LaH_x$, (x=10.2-11.1), synthesized after laser heating at pressures above 160 GPa. Strikingly, they found hydrogen to be in a highly diffusive state at room temperature. Over several weeks, dehydrogenation drove the composition toward that of the precursor material $LaH_3$. Even more surprisingly, the $LaH_x$ samples with x ranging from near 10 to 3, stored for 6-52 days, exhibit similar resistive transitions at nearly identical $T_c$ values of 80 K and all reach a zero-resistance state (Fig. 3k, 3l) [21].

Finally, we would like to point out that a sharp transition into zero resistance is experimentally difficult to observe in partially superconducting mixed-phase samples, whereas it appears in most inhomogeneous superhydride samples under high pressure. The many similarities in zero resistance transitions among the materials discussed in this review raise the question of whether they are coincidental or reflect a common mechanism.

## Acknowledgments

Z.R. is grateful for helpful discussions with several colleagues. This work was supported by the National Natural Science Foundation of China (Grants No. 12074414), the National Key Research and Development Program of China (Grant No. 2021YFA1401800), the Strategic Priority Research Program of Chinese Academy of Sciences (Grant No. XDB25000000), and the CAS Superconducting Research Project (Grant No. SCZX-0103).